\documentclass[useAMS,usenatbib]{mn2e}
\usepackage{graphicx}
\pdfoutput=1

\title[Reflected light from 3D exoplanetary atmospheres and simulation of HD 209458b]
{Reflected Light from Three Dimensional Exoplanetary Atmospheres and Simulation of HD 209458b}
\author[B. Hood et al.]{Ben Hood$^{1,}$$^2$, Kenneth Wood$^1$,  
Sara Seager$^2$\footnote{Current mailing address: Department of Earth, Atmospheric, and Planetary Sciences, Massachusetts Institute of Technology, Cambridge, MA 02139, USA}, Andrew Collier Cameron$^1$
\newauthor 
\\
$^1$School of Physics \& Astronomy, University of St Andrews, North Haugh,
St Andrews, Fife KY16 9SS, Scotland\\
$^2$Department of Terrestrial Magnetism, Carnegie Institution of Washington
5241 Broadbranch Road, Washington, DC 20015, USA\\
}

\begin{document}

\maketitle

\begin{abstract}

We present radiation transfer models that demonstrate that reflected light levels from three dimensional (3D) exoplanetary atmospheres can be more than 50\% lower than those predicted by models of homogeneous or smooth atmospheres. Compared to smooth models, 3D atmospheres enable starlight to penetrate to larger depths resulting in a decreased probability for the photons to scatter back out of the atmosphere before being absorbed. The increased depth of penetration of starlight in a 3D medium is a well known result from theoretical studies of molecular clouds and planetary atmospheres. For the first time we study the reflectivity of 3D atmospheres as a possible explanation for the apparent low geometric albedos inferred for extrasolar planetary atmospheres. Our models indicate that 3D atmospheric structure may be an important contributing factor to the non-detections of scattered light from exoplanetary atmospheres. We investigate the self-shadowing radiation transfer effects of patchy cloud cover in 3D scattered light simulations of the atmosphere of HD209458b. We find that, for a generic planet, geometric albedos can be as high as 0.45 in some limited situations, but that in general the geometric albedo is much lower. We conclude with some explanations on why extrasolar planets are likely dark at optical wavelengths.

\end{abstract}


\section{Introduction}
Since the discovery of the planet orbiting 51~Peg over a decade ago \citep{may95}, the detection and characterization of extrasolar planets has become an area of intense study. To date, over two hundred and seventy extrasolar planets have been discovered primarily by the measurement of the doppler shifting of stellar spectral lines due to the motion of the star around the centre of mass of the exoplanetary system. Other indirect techniques that have succesfully detected extrasolar planets include transit searches and microlensing (e.g. \cite{pon06,bea06}). The first detection of direct light from an extrasolar planet was the infrared emission due to reprocessed starlight in the warm atmospheres of the planets orbiting TrES-1 and HD 209458 \citep{cha05,dem05}. Additionally, a handful of planets may have been detected using direct imaging \citep{neu05}

Many extrasolar planets orbit within a few astronomical units of their parent star. The direct detection of reflected starlight from exoplanetary atmospheres is therefore a realistic goal (e.g., \citet{sea00}). However, despite intense effort observers have been unable to detect stellar light reflected from extrasolar planetary atmospheres. \citet{cam99} and \citet{cha99} developed techniques that exploit the doppler effect to separate the spectral lines of an extrasolar planet from the lines of the parent star, but both have been unable to definitively detect the faint signature. A detection of reflected light from the planet orbiting Tau Bootis by \citet{cam99} was unverifiable in the following observing season \citep{cam04}. Subsequent work has set an upper limit of 0.39 for the planet's geometric albedo, ($A_g$), assuming a grey albedo and Venus-like phase function \citep{lei03}. Using similar methods the group found the geometric albedo of the companion to HD 75289 has an upper limit of $A_g < 0.16$ \citep{lei03a}. Analysis of data from the {\it MOST} satellite has set an optical (400-700 nm) limit of $A_g < 0.17$ on HD~209458b \citep{row07}.

These non-detections of reflected light and upper limits on the geometric albedo imply that compared to solar system planets, the extrasolar planets are very dark at optical and near-infrared wavelengths. In comparison, the geometric albedos of solar system gas giant planets range from 0.41 to 0.52 \citep{cox00}. Particles in the exoplanetary atmospheres could have very low scattering albedos or the particles could have scattering phase functions that are very forward throwing so that starlight is directed deep into the atmosphere from where it has a small probability of scattering back out. \cite{sud00} explored parameters which can influence the geometric albedo of an extrasolar planet, with a one-dimensional radiation transfer model. The focus of our paper is the effect of three-dimensional atmospheric structure on the radiation transfer of starlight and hence the overall reflectivity of exoplanetary atmospheres.

Previous studies have shown the importance of three-dimensional radiation transfer effects on the penetration of starlight into dark clouds \citep{boi90}, the analysis of reflection nebulae \citep{mat02}, and the infrared spectrum from dusty ultracompact H~{\sc ii} regions \citep{ind06}.  In the planetary atmosphere community much effort has been devoted to developing three-dimensional radiation transfer techniques to study the penetration of Solar radiation through clouds in the Earth's atmosphere \citep{cah05}. Studies of clouds in the Earth's atmosphere have shown a fractal structure \cite{lov82}. Additionally, \citet{kuc04} showed that a three-dimesional Monte Carlo radiation transfer simulation in fractal clouds results in a more accurate representation of radiation penetration in real clouds than a plane-parallel homogenous simulation. Though we lack the large amount of data of atmospheric scientists studying the Earth, this research is very relevant to extrasolar planetary atmospheric simulations.

The work we present in this paper is motivated by the non-detections of reflected light from exoplanets and the clear three-dimensional nature of clouds in the atmospheres of the Earth and other planets in the Solar System. We present three-dimensional\footnote{We use 3D to mean that we can vary the density and characteristics of our model for each point in our x,y, and z dimensional grid, and introduce structure that varies in every direction.} radiation transfer simulations of the reflected light from inhomogeneous exoplanetary atmospheres. We have not computed the 3D temperature and pressure structure of an atmosphere, rather to investigate 3D radiation transfer effects on reflected light we take 1D atmospheres and convert them to 3D via a hierarchical clumping algorithm. This approach allows us to identify and characterize the different reflectivities of uniform and 3D atmospheres.  We then extend our analysis to study a more detailed atmospheric model incorporating multiple species of scatterers, cloud decks, and vertical density gradients within the atmosphere. Although we do not solve for the 3D temperature, pressure, and density structure, we attempt to build a more realistic simulation and will focus specifically on current atmospheric models for the extrasolar planet HD 209458b. We have simulated the atmosphere and found that in all cases the geometric albedos for the simulations of extrasolar planet HD 209458b are lower, and in many cases much lower, than that of Jupiter, suggesting radiation transfer effects in a 3D atmosphere might help explain why to date, reflected light has not been detected from an extrasolar planet. 

In \S\ref{sec:code} we describe our numerical radiation transfer code, the algorithm we use for generating a hierarchically clumped 3D atmosphere, and the adopted scattering properties of particles within the atmosphere. Results of our radiation transfer simulations for a wide range of atmospheric optical depth, porosity, and particle scattering properties are presented in \S\ref{sec:simres} and \S\ref{sec:dis} summarizes our generalized results with respect to the scattered light levels. \S\ref{sec:209458} presents 3D radiation transfer models for the scattered light from the atmosphere of  HD 209458b and in \S\ref{sec:209sum} we summarize our findings.

\section{Model Ingredients}
\label{sec:code}
This section describes the radiation transfer technique, atmospheric geometry, and particle scattering properties we adopt for our 3D radiation transfer simulations.

\subsection{Monte Carlo Radiation Transfer Code}
\label{sec:mcrt}
For our reflected light simulations we use a 3D Monte Carlo scattering code, described in \citet{woo99}. This code simulates radiation transfer through a 3D linear Cartesian grid onto which the density structure is discretized. The code accurately treats multiple, 
anisotropic scattering for any analytic or tabulated angular scattering phase function. We do not consider the reprocessing of absorbed stellar photons. This assumes that all photons that are absorbed by the planetary atmosphere are re-emitted at wavelengths that are well separated from the incident wavelength. Therefore our simulations will be accurate at optical and near-infrared wavelengths where reflected light dominates the planetary spectrum and thermal emission from the planetary atmosphere is negligible. To construct a simulation for a reflected spectrum using our technique we would run our code multiple times (wavelength by wavelength) and change the scattering properties of the medium to be those appropriate for each wavelength in the incident spectrum. 

The density grid we employ for the simulations in this paper has 200 cubical cells on a side, for a total of $8\times 10^6$ cells. Our atmosphere is semi-infinite, so that photon packets exiting the grid through $x$ or $y$ faces, re-enter the grid on the opposite face. Thus photons can only escape through the upper and lower $z$ boundaries of the grid.  See \S\ \ref{sec:lbc} below for our treatment of different lower $z$ boundary conditions. Because of our semi-infinite treatment of the atmospheric simulation, we are unable to reproduce the small amount of light which might otherwise pass through a tenuous atmosphere when the planet is directly between the star and the observer (at a 180\degr phase angle). Since we are considering reflected light and not transmission spectra (see \cite{sea00a}), this is not an obstacle to our approach.

\begin{figure}
  \includegraphics[angle=0,width=.5\textwidth]{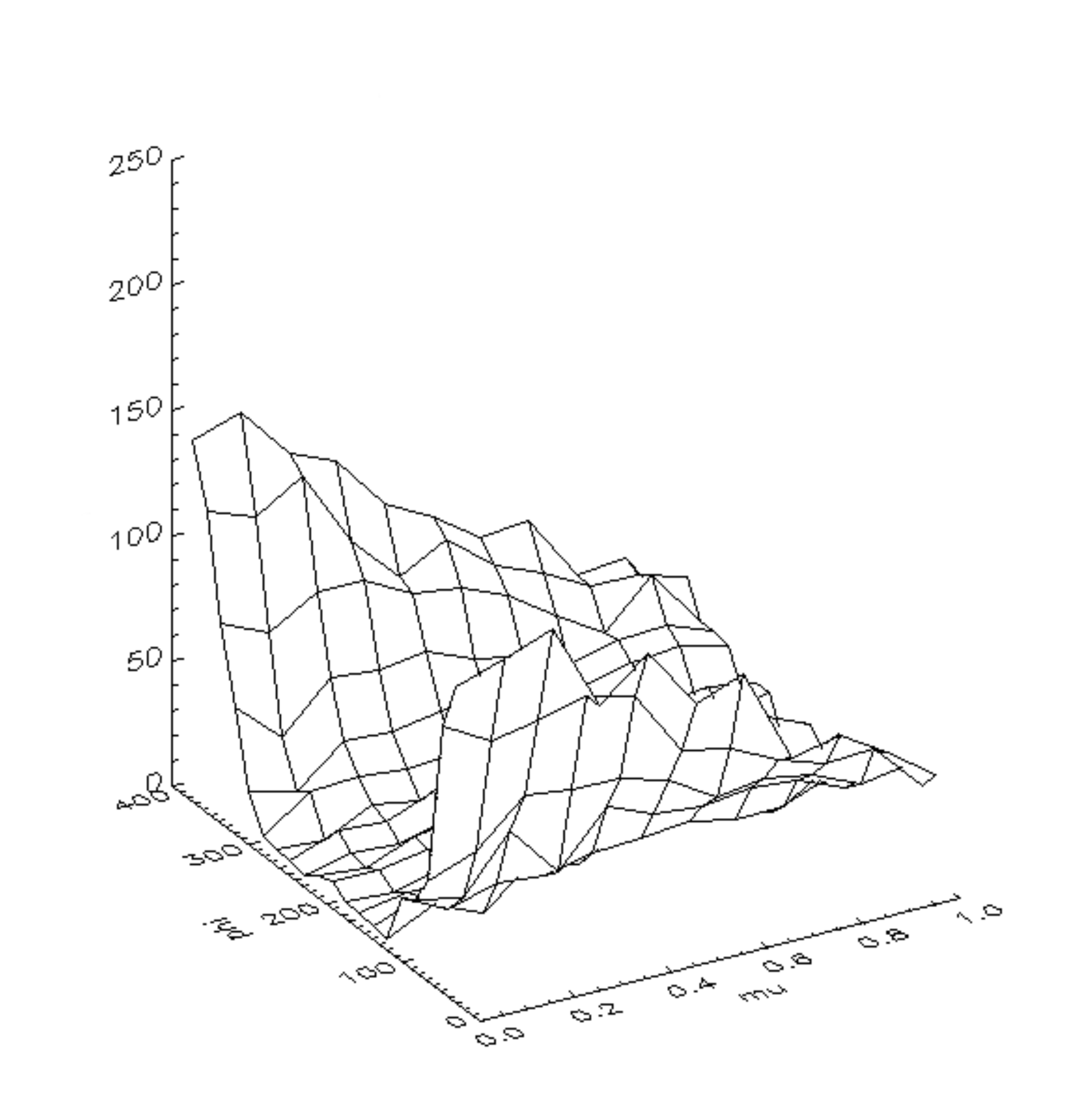}
  \caption{Reflected light from one of our semi-infinite simulations.  The 
density grid is uniformly illuminated by plane parallel rays at an angle of 
$20^\circ$ to the surface.  The reflected light is plotted as a function 
of longitude ($\phi$) and latitude ($\mu = \cos\theta$) around the grid, and is normalized to isotropic, non-absorptive scattering.}
  \label{fig:sha}
\end{figure}

The grid is illuminated from the outside with photons entering the upper $z$ face at specified angles representing the incident angle of stellar photons on the atmosphere. When photons exit the grid through either the upper or lower $z$ faces, they are binned in solid angle according to their direction of travel. The output of a simulation is the direction dependent reflected light levels in 400 evenly spaced solid angle bins (20 divisions in $\cos\theta$ and 20 divisions in $\phi$). We ran simulations for each realization of the atmosphere geometry with incident angles in the range $0^\circ$ to $85^\circ$ (measured from the normal to the upper $z$ face of the simulation grid) in $5^\circ$ increments. The results of those simulations are the flux emitted in each of the 400 solid angle bins, and can be presented graphically, as in figure \ref{fig:sha}  

Following other theoretical simulations of reflected light (e.g., \citet{sud05}), the results of our reflected light simulations from the semi-infinite models are tiled to generate the total reflected light from the planet. In the semi-infinite simulation we specify the incident angle of radiation and the output is the angle dependent reflectivity.  Using the results from the semi-infinite simulations we may then specify an incident angle of radiation, an emergent angle of scattered radiation and find the reflected light level compared to a Lambertian\footnote{A Lambertian surface is completely non-absorbing and scatters an equal intensity in all directions.} scattering surface. Using equation 9.8 from \citet{sob75}
\begin{equation}
\displaystyle
H(\alpha)=\int_{\alpha-\pi/2}^{\pi/2}\cos(\alpha-\omega) 
\cos \omega d\omega \int_0^{\pi/2}\rho(\eta,\zeta,\phi) \cos^3 \psi d\psi
\end{equation}
we determine $H(\alpha)$, the radiation flux emerging from a planet with respect to the planetary phase $\alpha$. In this equation $\eta$ is the cosine of the angle of emergence with respect to the outward normal, $\zeta$ is the cosine of the angle of incidence with respect to the outward normal, $\phi$ is the azimuthal angle between those two angles. The planetocentric coordinates of latitude and longitude are represented by $\psi$ and $\omega$. Finally, $\rho$ is the output from our Monte Carlo simulations and it is the reflection coefficient, or the flux scattered in a particular direction from the surface compared to an isotropically scattering surface. Our treatment of tiling results from many semi-infinite simulations follows that of \citet{sud05}, but we use Monte Carlo techniques for the 3D radiation transfer.

\subsection{3D Atmospheric Density Structure}
\label{sec:atmgeo}

The goal of this paper is to investigate the effects of 3D density structures and 3D radiation transfer on the reflectivity of extrasolar planets.  Therefore we begin with the simplest geometry of a uniform density atmosphere (no vertical stratification) and compare the resulting reflectivity with 3D atmospheres of the same total mass.  As mentioned in the introduction, we do not compute the 3D density, pressure, and temperature structure of an atmosphere, but convert smooth 1D atmospheric densities to 3D using the hierarchical clumping algorithm of \citet{elm97} as described in \citet{mat02} and modified in \citet{woo05}. 

The hierarchical clumping algorithm randomly casts $N_1$ points in 3D space at the first level. At the second level $N$ points are cast around each of the $N_1$ points from the first level, but with a separation in the range $\pm \Delta^{(1-H)}/2$, where $H$ is the hierarchical level and $f = \log N / \log \Delta$ is the fractal dimension of the hierarchical structure. At all subsequent levels a further $N$ points are cast around each of the points cast at the previous level, again with separation in the range  $\pm \Delta^{(1-H)}/2$.  The density in our Cartesian grid is proportional to the number of points in each cell that were cast at the final hierarchical level. See Figure~5 in \citet{woo05} for a 
graphical representation of the algorithm and how the density in each cell is determined.

In the simulations presented here we use five hierarchical levels, $N=32$, and fractal dimension $f= 2.6$.  This value of the fractal dimension approximately corresponds to a projected two-dimensional area-perimeter dimension of 1.36, appropriate for clouds in the Earth's atmosphere \citep{lov82}.

We convert the uniform density to 3D using this algorithm and leave a fraction of the mass, $f_{\rm smooth}$, in a smooth density component, the remainder being distributed according to the hierarchical algorithm.  The minimum vertical optical depth for a 3D model is $\tau^{3D}_z = f_{\rm smooth} \tau_z$, where $\tau_z$ is the optical depth of the smooth model and $f_{\rm smooth}$ is the mass fraction smoothly distributed in the 3D model. This optical depth occurs for sightlines along which there is no material distributed by the hierarchical clumping algorithm.  

To investigate different porosity levels of the atmosphere we adopt the approach of \citet{woo05} and vary the number of points, $N_1$, cast at the first hierarchical level. For small values of $N_1$ the porosity will be high and as $N_1$ increases the density approaches a more uniform distribution. Specifically we investigate atmosphere models with $N_1 = 2$, 8, 32, and 128. These values correspond to porosity values of around 99\%, 95\%, 80\%, and 45\% respectively, where we define porosity as the fraction of the volume occupied by the smooth density component. We also determine the covering factor of the hierarchically distributed clumps in the grid by creating column density maps 
viewed along the $z$ axis. For the $N_1$ values used in the paper the covering fractions are approximately 15\%, 45\%, 90\%, and 100\% respectively. These values of the porosity and cloud coverage do not change significantly (within a few percent) for different random castings of the first $N_1$ points in the hierarchical clumping algorithm. For the rest of the paper we refer to different 3D simulations according to their porosity or covering factors. In the paper we refer our 3D results to ``the equivalent smooth model," which means the smooth model with a particular optical depth (i.e., uniform density or mass) which is then converted to 3D using the hierarchical clumping algorithm.

Figures~\ref{fig:frac} and \ref{fig:vox} show examples of the hierarchical density grid. Figure~\ref{fig:frac} shows column density maps for the four different porosity values we investigated.  The lowest column density in each map arises when there are no hierarchically distributed clumps along the line of sight, and so in these maps black is not zero density, but the column density due to the smooth component only in the density grid. Figure~\ref{fig:vox} shows a voxel projection of the density grid for a porosity value of 80\%. Note that our implementation of a semi-infinite grid for the radiation transfer simulation is equivalent to an infinite sequence in $x$ and $y$ of such density cubes.

We have the ability to change our grid size in order to investigate finer or coarser resolution, but with current settings our individual simulations represent approximately a 1200 x 1200 x 1200 km cube on the surface of a Jupiter-sized planet. When we simulate HD209458 in \S\ \ref{sec:209458}, we adjust our simulation to investigate a deeper portion of the atmosphere with a grid of $8000 \times 1600 \times 1600$ km.

\subsection{Particle Scattering Properties}

The Monte Carlo code simulates the scattering of stellar photons off particles in the planetary atmosphere. The code can treat analytic or tabulated scattering phase functions. In our initial investigation we represent the angular shape of scattering with the Henyey-Greenstein phase function \citet{hen41} characterized by the single parameter, $g$,
\begin{equation}
HG(\upsilon) = {1\over {2}}{{1-g^2}\over{(1+g^2 - 2g\cos\upsilon)^{3/2}}}
\end{equation}
where $\upsilon$ is the scattering angle. The scattering asymmetry parameter controls the shape of the phase function, with $g=0$ giving isotropic scattering, positive $g$ for forward throwing, and negative $g$ for a phase function dominated by back-scattering.  

The scattering albedo $a$, determines the fate of photons at their interaction locations in the atmosphere. Absorbed photons are terminated and assumed to be reprocessed to long wavelengths and do not contribute any flux at the simulation wavelength.

Typical values for the albedo and scattering asymmetry parameters in the optical are $a \sim 0.5$ and $g \sim 0.5$, based on Mie theory calculations for grain types assumed to be present in extrasolar planetary atmospheres. Enstatite, iron, and corundum grains are expected in close-in giant planet atmospheres \citep{sea00}. 

\subsection{Lower Boundary Conditions}
\label{sec:lbc}

Simulated photons are injected into the density grid as described above and may exit the grid through the upper $z$ face in which case they are placed into direction-of-observation solid angle bins. For photons that exit the simulation through the lower $z$ face we investigate two scenarios, absorption or Lambert reflection. In the case of absorption, this may represent the absorption of photons by an optically thick layer below the cloud deck. The second case is equivalent to photons being isotropically scattered off a planetary surface; the simulation returns the light curve expected for a Lambert sphere when the optical depth of the atmosphere is zero (see \S\ \ref{sec:tes} below). These two 
boundary conditions have the largest influence on the reflected light for atmospheres where photons penetrate to the lower $z$ boundary, such as optically thin atmospheres and those with very low porosity.

\begin{figure}
  \includegraphics[angle=0,width=.5\textwidth]{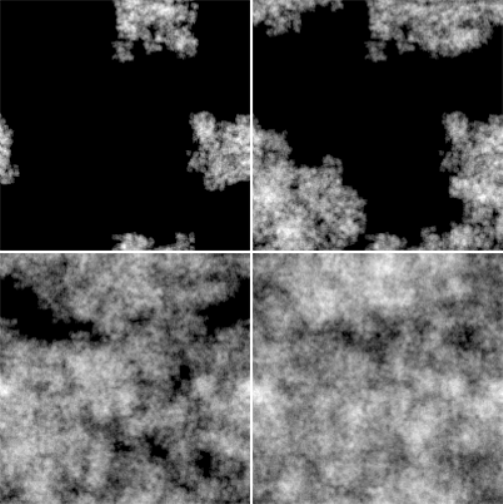}
  \caption{Vertical column density maps for four different porosity levels produced by our hierarchical clumping algorithm.  We define porosity to be the fraction of our simulation grid that contains only the smooth density component.  
The figures show porosity levels of 99\% (top left), 95\% (top right), 
80\% (bottom left), and 45\% (bottom right).  }
  \label{fig:frac}
\end{figure}

\begin{figure}
  \includegraphics[angle=0,width=.5\textwidth]{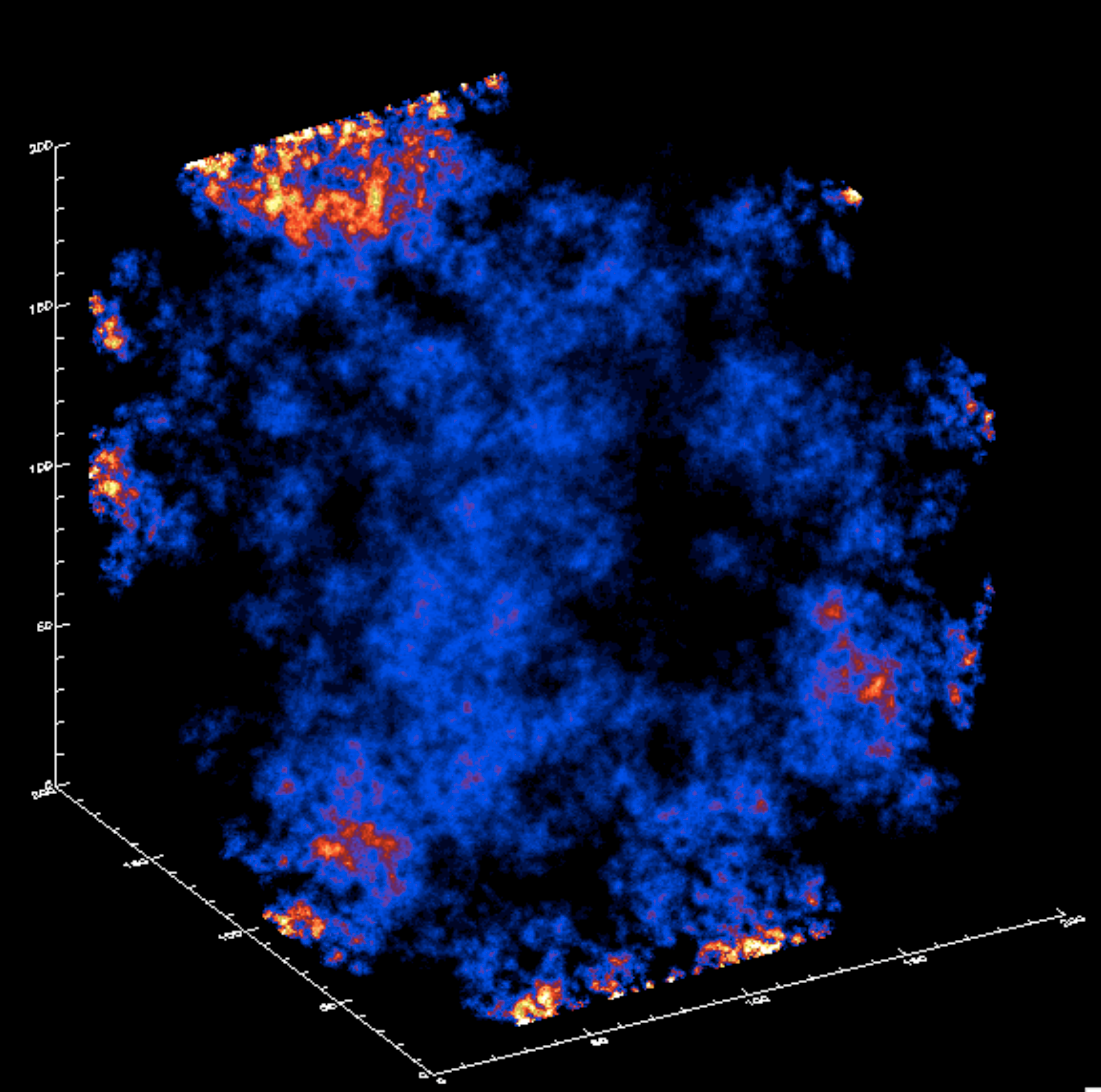}
  \caption{A voxel projection of the 3D density distribution for a density grid 
with porosity of 80\%, corresponding to the lower left panel in Fig.~\ref{fig:frac}.}
  \label{fig:vox}
\end{figure}

\subsection{Albedo definitions}
We use two definitions of albedo throughout this paper. The first albedo, called single-scattering albedo, refers to the albedo 
of the particles in the atmosphere. We assign an albedo to each scattering species. For instance, if we assigned an albedo of 0.6 to a 
scattering species representing iron, then approximately 60\% of the Monte Carlo photons interacting with iron will be scattered, and 40\% absorbed. 

The second albedo we use is the geometric albedo, $A_g$, which describes the albedo of the entire planet. The geometric albedo is defined relative to the amount of light reflected from a flat, Lambertian disk with the same radius as the planet, at the same distance from the star as the planet. Comparing geometric albedo to the Lambert result is only strictly correct when the planet is at a phase angle of zero degrees. However, the reflected light at the other phase angles can be compared to this zero phase geometric albedo and we present this in our figures below.

\subsection{Code Testing}
\label{sec:tes}
We have tested our Monte Carlo scattering codes against analytic solutions and other independently developed radiation transfer codes. For scattering simulations, Chandrasekhar's solution for a plane parallel Rayleigh scattering atmosphere illuminated from below is one of the standard benchmarks. Our codes accurately reproduce Chandrasekhar's results for the angular dependence of the emergent radiation (intensity and polarization) from the top of the atmosphere (e.g., see \citet{woo96}). In this paper we wish to tile results from a semi-infinite Monte Carlo simulation to generate a phase function for the reflected light from the planet as a whole. To test that we correctly implement the tiling described in \S\ \ref{sec:mcrt}, we set up a simulation to represent a Lambert sphere. In this simulation the atmosphere has zero optical depth and photons at the lower boundary reflect isotropically.  

Our tiling procedure reproduces the Lambert results to within 1\% at small phase angles, and to much better accuracy at larger phases. These small discrepancies are due to the coarseness of our numerical integration. Increasing the angular resolution of the integration by a factor of ten reproduces the Lambert phase function to within 0.01\%.  However, the small increase in accuracy does not justify the significant increase in computational time, so the results presented in the rest of the paper use the lower angular resolution.

\subsection{Varying the cloud distribution for the same porosity}
\label{sec:cloud}
To set up the 3D density structures we use an initial random number as a seed for the hierarchical clumping algorithm. Different random seeds  produce different arrangements of clouds for the same porosity value and most likely will yield different levels of reflected light. We have investigated the variations in reflectivity arising from different 3D structures with the same porosity levels and find the largest 
differences (up to 20\%) occur for simulations with very high porosity. In these situations there are few cloud complexes in the simulation grid and if clouds lie above one another the covering factors can be very different for the same porosity level. Smoother atmospheres exhibit much smaller variations (typically 5\%) among different cloud geometries with the same porosity.

\section{3D Scattered Light Simulations}
\label{sec:simres}
\subsection{Fiducial Smooth Model}
We have chosen a set of uniform density fiducial models against which we compare reflected light levels with those from the 3D models. 
The fiducial models have a uniform density, phase function asymmetry parameter $g=0.5$, particle scattering albedo $a=0.5$, and total vertical optical depths through the grid of $\tau_z = 0.1$, 1, 10, and 100. The optical depths and scattering parameters of the fiducial models are similar to those expected in extrasolar planetary atmospheres \citep{sea00} and provide a benchmark against which we can explore the effects of 3D density structures. The lower boundary condition for these fiducial models are that all photons at the lower boundary are absorbed. 

\begin{figure}
  \includegraphics[angle=0,width=.5\textwidth]{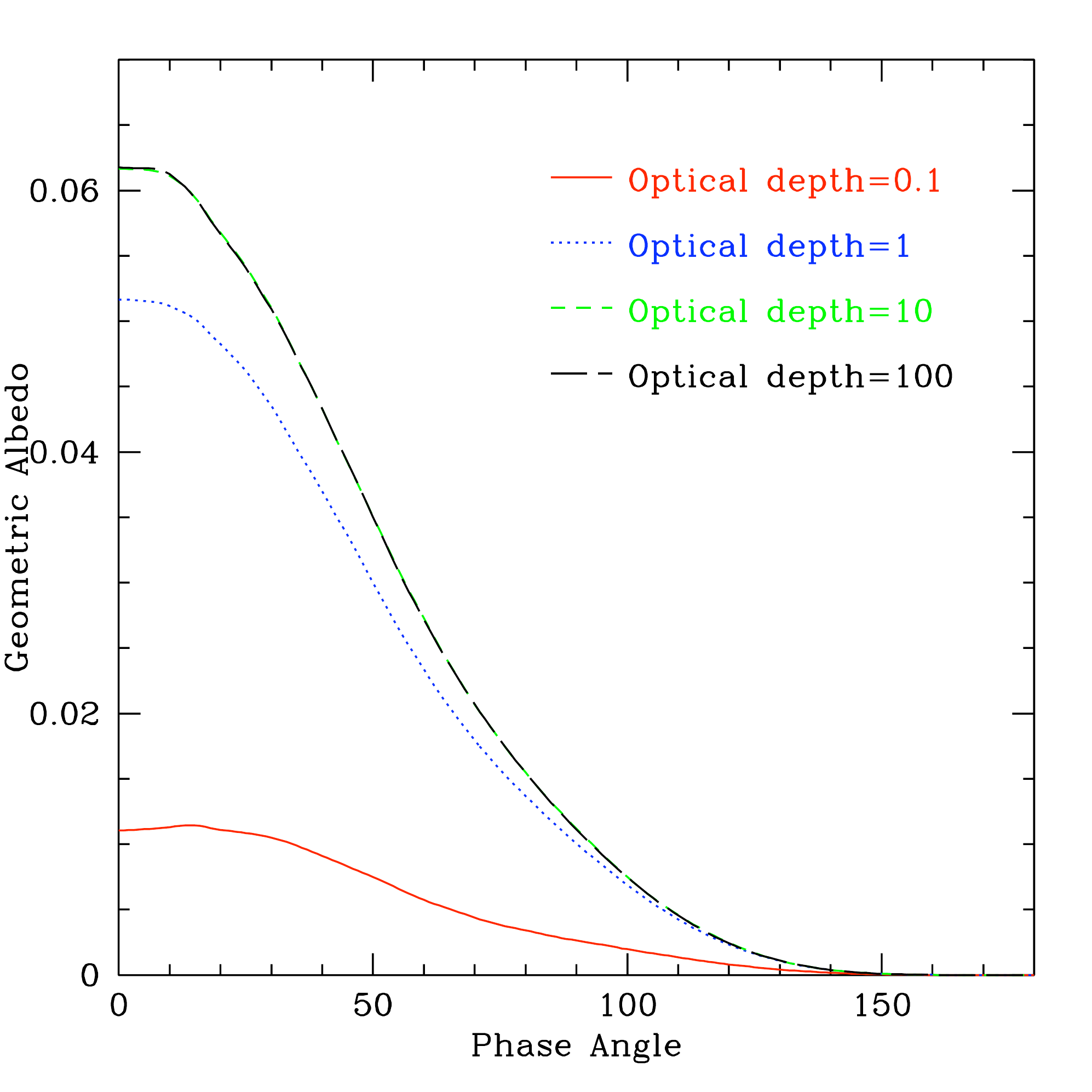}
  \caption{Effect of $\tau_z$ on reflectivity for our set of fiducial models at a range of optical depths. All models have single particle albedo $a=0.5$, an absorbing lower boundary, and an atmosphere with a uniform density structure.  For $\tau>10$ the results do not change.
}
  \label{fig:fid}
\end{figure}

Figure \ref{fig:fid} shows the derived geometric albedos as a function of phase angle for the tiled simulations for these four smooth models. An immediate result from the fiducial models is that the reflectivity and hence geometric albedo does not change for optical depths greater than $\tau_z=10$.  At high optical depths the reflectance is due to scattered photons that only penetrate to a small physical depth in the atmosphere. The geometric albedo is dominated by photons that are scattered off the top of the optically thick atmosphere.

\subsection{Parameter Exploration}
\label{sec:par}
The main focus of our paper is to investigate the effects on reflected light levels of 3D radiation transfer in clumpy atmospheres. However, as has been reported in other theoretical reflected light studies of 1D atmospheres, many parameters can affect the overall reflected light levels \citep{sud00,bar05,sud05}. As described in detail in \citet{bah07}, we have explored a very wide parameter space and investigated 
the effects of varying particle albedo, phase function, and lower boundary conditions. In this section, we first briefly describe these results before presenting figures for our main study: the effects of atmospheric porosity on reflected light levels from extrasolar planets.

\subsubsection{Particle albedo and phase function}
One of the most significant parameters set within the simulation is the albedo of the scattering particles in the atmosphere. This was set to $a=0.5$ in the fiducial models in Figure \ref{fig:fid} and produced a geometric albedo for the planet of $A_g \sim 0.06$ at small phase angles. Increasing the particle albedo to $a=0.99$ dramatically increases the planet's geometric albedo to $A_g = 0.55$ at small phase angles. 
Particle albedos of $a=0.8$ and $a=0.2$ produce geometric albedos at zero phase of $A_g = 0.25$ and 0.03 respectively. Our results are in agreement with the geometric albedos calculated by \citet{dlu74} who numerically solve for the intensity of radiation emerging from an atmosphere using iterative techniques. 

The single scattering phase function determines the angular distribution of scattered photons. In our initial investigations we use the Henyey-Greenstein phase function described by the asymmetry parameter $g$. A strongly forward scattering phase function will scatter incident stellar photons deeper into the atmosphere with the effect of reducing the overall geometric albedo of the planet. Specifically we find zero phase geometric albedos can be reduced by a factor of three when the asymmetry parameter is varied from isotropic scattering ($g=0$) to very forward scattering ($g=0.9$).

\subsubsection{Lower boundary condition}
The bottom layer of our simulation represents the bottom of the atmosphere. On a rocky planet, the bottom layer should act like a surface of a rocky planet, be it snow, water, or sand. On a gas giant planet, the bottom layer is only defined as the limit of our simulation. We must specify how much atmosphere to simulate, and how to treat the photon packets that make it through the entire (simulated) atmosphere. We assign one of two characteristics to the bottom layer: Lambert or absorptive. A Lambert surface isotropically reflects all radiation, while an absorptive surface absorbs all the incident photons. Actual absorbed photons are often reprocessed to longer wavelengths and do not contribute to the optical scattered light.

Varying the bottom layer makes a huge impact on a small subset of our models with sparse cloud coverage or very low optical depths. In these situations the Lambert surface produces much larger geometric albedos than the absorbing lower boundary condition. For atmospheres with optical depths $>10$ or porosity levels $<80\%$, the lower boundary does not affect our results since the vast majority of stellar photons interact with the upper levels of the atmosphere and do not penetrate to the bottom.

\subsection{Optical Depth effects in smooth and 3D atmospheres}
\label{sec:opd}

\begin{figure*}
\begin{center}
  \includegraphics[angle=0,width=\textwidth]{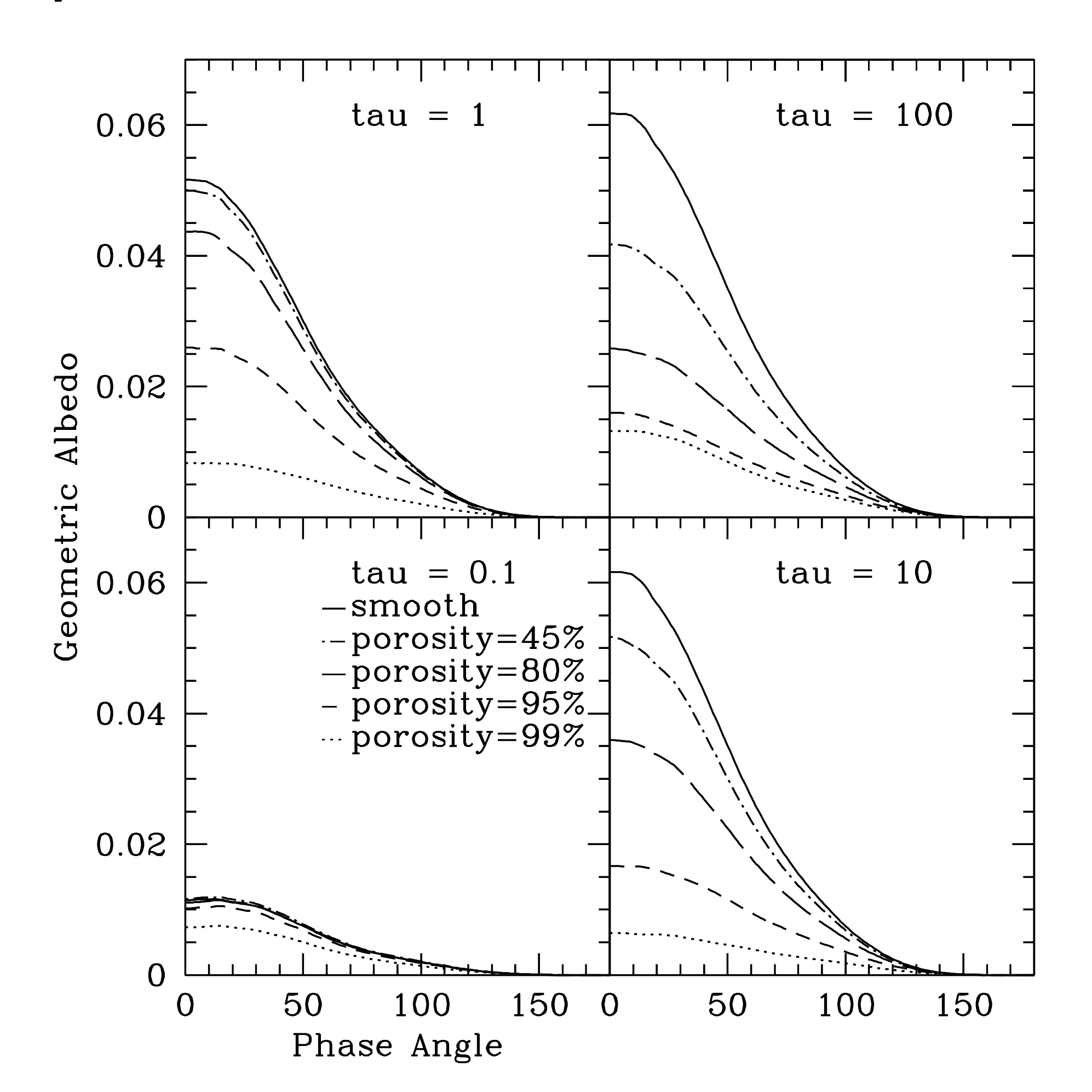}
  \caption{Reflected light phase functions for our set of 3D atmospheric models with 
different porosity levels.  All models here used $a=0.5$, $g=0.5$, absorbing bottom 
layer, $f_{\rm smooth} = 0$.  Notice that 3D effects lead to a reduction in the 
reflectivity for {\it all} optical depths considered (the possible exception being $\tau = 0.1$, where the results of the smooth model are actually indistinguishable from several porous models, due to photon noise at that low level.) }
  \label{fig:resfrac}
  \end{center}
\end{figure*}

Figure \ref{fig:resfrac} presents results of simulations for four different atmospheric porosity levels and four different optical depths of the equivalent smooth model. Along with the smooth model, this figure therefore shows twenty different reflected light models covering what we believe to be a range of optical depth and porosity relevant to atmospheric cloud structures.

The atmospheric optical depth determines photon penetration and is a major factor in determining a planet's geometric albedo. With a small optical depth, photon packets pass through the atmosphere with no interaction and the lower boundary condition determines the reflectivity. 

As the optical depth increases, the 3D models and the smooth models react differently. The geometric albedos for smooth and 3D models initially increase with optical depth. Above $\tau_z \sim 10$ the smooth models maintain a constant geometric albedo. However, for 3D atmospheres the geometric albedo begins to decrease for high optical depths. This is a result of the increased depth of penetration of photons in a 3D atmosphere compared to the equivalent smooth model. While photons that penetrate deep into the atmosphere may scatter back out via low optical depth paths, there is also a high probability that they will be absorbed by optically thick clumps encountered along their random path back out of the atmosphere. Recall from \S\ \ref{sec:mcrt} that the equivalent smooth model has the same total mass as a 3D model, but in the 3D model the mass is redistributed via the hierarchical clumping algorithm resulting in regions of significantly lower and higher optical depths and thus an overall increase in depth of penetration. This is a well known result from three-dimensional radiation transfer (e.g., \citet{boi90}).

In addition to $\tau_z$, figure \ref{fig:resfrac} shows that the atmospheric porosity has a significant impact on a planet's reflectivity. As described in \S\ \ref{sec:cloud}, we vary the porosity of the 3D models and a high porosity results in only a few dense cloud structures in the grid. For lower porosities the hierarchically distributed mass results in a more continuous cloud structure and larger cloud covering factor (see Figure \ref{fig:frac}).

The main trends that we observe in Figure \ref{fig:resfrac} are 1) greater cloud coverage results in a higher overall albedo, 2) compared to the equivalent smooth models, 3D atmospheres have a lower $a_g$, and 3) different porosities have different optical depths beyond which the geometric albedo ceases to rise and begins to decrease.

Intuitively, if more of the atmosphere is covered with clouds, then the overall albedo will be higher. For an optical depth $\tau=10$, the simulation with 100\% cloud coverage has a zero phase geometric albedo $A_g = 0.08$; while cloud coverages of 90\%, 45\%, 
and 15\% give zero phase geometric albedos $A_g = 0.06$, 0.03, and 0.01 respectively. This is expected because more cloud coverage will result in more photons being reflected from the upper layers of the atmosphere. Note these simulations have an opaque lower boundary and $f_{\rm smooth}=0$. As $f_{\rm smooth}$ is increased the geometric albedo increases towards the values for the smooth models.

When compared to the equivalent smooth models, all our 3D simulations with significant optical depth and internal structure have lower geometric albedos. An albedo bias between uniform and one-dimensionally fractal models has been reported in atmospheric science literature, specifically modeling the Earth's atmosphere \citep{cah94,cah94a}. If treated as plane-parallel, California stratocumulus clouds have a relative albedo bias of 15\%, meaning that smooth simulations will suggest an albedo 15\% higher than the albedo of the actual clouds, which have a great deal of fractal structure. 

\subsection{Discussion of Scattered Light Levels}
\label{sec:dis}

Our investigation finds that 3D radiation transfer effects in an inhomogeneous atmosphere lead to lower levels of reflected light compared to uniform atmosphere models of the same total mass. For inhomogeneous atmospheres low density paths allow the photons to penetrate deep into the atmosphere where there is a higher probability they will be absorbed in a dense cloud on their path back up through the atmosphere. This lowers the geometric albedo and is a 3D radiation transfer effect. However, perhaps the most important single parameter is the albedo of the individual scattering particles in the atmosphere \citep{sud05,bar05,sea05}. Very high albedos associated with enstatite and iron particles (typically $a>0.9$, see next sections) can result in reflected light levels an order of magnitude higher than an atmosphere where the particles have, for example, $a=0.5$. 

Overall, the geometric albedos are much lower than expected and make reflected light detection very difficult. Other papers have specifically discussed the low geometric albedos of extrasolar planetary atmospheres, including \citet{mar99}, \citet{sea00}, while \citet{dlu74} found low theoretical albedos for generic planets. Marley mentions that, using their model, planets warmer than 400 K will be dark and that most planets will be dark in reflected light beyond a 600 nm wavelength. Seager shows that at wavelength 480 nm the geometric albedo of her model varies between below 0.01 and 0.4, depending on the mean particle size. 

Finally, \citet{dlu74} used an analytical method of calculating the intensity of radiation from a homogeneous atmosphere to find that isotropic scattering in an atmosphere with particles with a single scattering albedo of 0.9 would only result in a geometric albedo of 0.3. Their models specifically illustrate the necessity of a very high single scattering albedo to produce a high geometric albedo. In a forward throwing atmosphere, a seemingly minute change in the single scattering albedo from 0.99 to 0.98 can change the geometric albedo from 0.34 to 0.26 \citep{dlu74}. It is possible that many researchers were expecting (or hoping) extrasolar giant planets would have the same high geometric albedos as Jupiter ($A_g=0.52$) and Saturn ($A_g=0.47$). However, at the temperatures of the extrasolar planets, there seem to be precious few condensates with the high single scattering albedos. This is a sharp contrast to the condensates of the gas giants of our own solar system, which typically have single scattering albedos of 0.996 \citep{cha97}. Much as we were expecting to find Jupiters around other stars, and instead found close-in giant planets, we were expecting to find Jupiter-like geometric albedos and instead are finding dark, low geometric albedos.

\section{Reflected Light from HD 209458b}
\label{sec:209458}

In the previous section we explained how inhomogeneous planetary atmospheres lead to very low geometric albedos compared to smooth atmospheres. In this section we construct a 3D scattered light model for HD 209458b, arguably the best studied extrasolar planetary atmosphere \citep{sea00a,sea03,dem05,sho08}. The goal of this model is to investigate 3D radiation transfer effects on the reflected light levels, in particular the geometric albedo limit from the MOST satellite. In addition, using recent 1D atmospheric models, we investigate limits on the atmospheric parameters such as cloud height, location, and composition. As with the models presented earlier, we do not construct a self-consistent 3D temperature-pressure-density model for HD209458b, but take a 1D atmosphere model and convert it to 3D using the hierarchical clumping algorithm described earlier. There have been recent 3D atmosphere models for HD209458b, but these models did not include 3D radiation transport. Therefore, what we present in this paper are the first 3D scattered light models for extrasolar planets and may be seen as a first step towards incorporating accurate 3D radiation transfer into global 3D atmospheric models.

For our HD209458b models we extended the Monte Carlo simulations to include exponential density gradients, multiple opacity sources mixed throughout the simulation grid, and scattering from pre-tabulated phase functions as may be computed from Mie scattering codes. In addition, we can limit the lower and upper boundaries of a cloud of scattering particles to specific heights in the atmosphere. 

In this section we detail how we determined our fiducial choices for the parameters for individual scattering species. We explain which parameters might reasonably be varied, and we simulate those variations to compare the geometric albedo of the varied atmospheres. We present results for scattered light at 500nm, at which wavelength absorbed photons will not primarily contribute to the optical flux. 

\subsection{HD209458b atmospheric parameters}
\label{sec:atmpar}
We must first compute the specific parameters of the simulations, such as the densities and opacities of the condensates. Two of the most important condensates in HD 209458b's atmosphere are believed to be enstatite ($\rm{MgSiO_3}$), with an albedo of 0.999 and iron (Fe), with an albedo of 0.685 \cite{ack01}. They both condense high in the atmosphere, which makes them significant contributors to the geometric albedo. Further, the lack of other high albedo species (in the temperature's and pressure's of HD 209458b) indicates that enstatite must be present in very reflective planets. Our focus in simulating HD 209458b will be on simulating a Rayleigh scattering atmosphere mass-dominated by hydrogen, with Mie-scattering clouds of both enstatite and iron. In all cases we will be simulating photon packets at 500 nm, and choose the optical properties (opacity, scattering, and phase function) accordingly. This is a reasonable choice of simulated wavelength due to HD 209458 being a G0V star.

\subsubsection{Density profile}
The smooth density structure we initially adopt is provided by the 1D atmosphere code described by \cite{sea05} which self consistently solves the equations of radiation transfer, radiative equilibrium, and hydrostatic equilibrium. (We should mention that there is some indication that HD 209458b is not in hydrostatic equilibrium at the top of the atmosphere, and this is a limitation of our simulations \citep{vid04}.) Using the temperature-pressure profile from Seager's code we compute the vertical hydrogen density structure assuming the ideal gas law holds throughout the atmosphere. It is this 1D density profile that we will convert to 3D using our hierarchical clumping algorithm. 

We find the mass density profile of hydrogen computed from Seager's code fits an exponential of the form,
\begin{equation}
\displaystyle
\rho= 3.4\times 10^{-8} e^{{-z}/{H}}\; {\rm{g}}\,{\rm{cm^{-3}}}
\label{eq:expatm}
\end{equation}
where $\rho$ is the density, $z$ is the height in cm in the atmosphere, and the scaleheight $H=1.04\times10^8$cm. The total height of the atmospheric simulation is determined from hydrostatic equilibrium to be $8.2 \times 10^8$cm, about 9\% of the radius of the planet.

Assuming solar abundances for the elements Mg, Si, and O, we construct the smooth 1D density structures of enstatite and iron. For every $10^{12}$ atoms of hydrogen, there are $10^{8.93}$ atoms of O, $10^{7.58}$ atoms of Mg, and $10^{7.55}$ atoms of Si, so the Si is our limiting factor \citep{cox00}. The density, $\rho$, which we computed above is the density of $\rm{H}_2$, which has a molecular weight of 2.02 ${\rm{g}}\,{\rm{mole}^{-1}}$. 

Once we have the moles of H, we can determine the moles of $\rm{MgSiO_3}$, since using the above ratios, we know that for every one H atom we \emph{can} have $3.548\times 10^{-5}$ molecules of $\rm{MgSiO_3}$. We also have a free parameter, $c$, which we will use to describe the percentage of enstatite that has condensed into the cloud. Thus, our density structure of enstatite condensed in the atmosphere is:

\begin{equation}
\displaystyle
\rho_{\rm{MgSiO_3}}=c \times \rho_H\times 3.59 \times 10^{-3}
\end{equation}

Similarly, we calculate the density of iron in the atmosphere assuming solar abundances. For every $10^{12}$ atoms of H, there are $10^{7.54}$ atoms of Fe \citep{cox00}. Using this ratio, for every H atom we will have $3.467\times10^{-5}$ atoms of Fe giving,

\begin{equation}
\displaystyle
\rho_{\rm{Fe}}=c \times \rho_H\times 1.917 \times 10^{-3}
\end{equation}
where again $c$ is the condensation fraction.

As noted in \S\ \ref{sec:atmgeo}, we previously made individual simulations in cubes, then we tiled those cubes onto the surface of a sphere in order to represent the entire planet. However, in order to match the height of the simulated data, we have had to remake our individual simulations into a rectangular grid, with $z$ being five times the length of $x$ or $y$. We maintained the cubic dimensions of our individual grid cells, and instead simply use five times as many grid cells in the $z$ direction (500 $z$ cells instead of 100 cells in $x$ and $y$). The height of our entire atmospheric simulation is $8.235\times10^8$ cm, and our individual grid cells are $1.647\times10^6$ cm on each side. Each of the cubic cells has a volume of $4.467\times10^{18} \rm{cm}^3$.

\subsubsection{Location of enstatite and iron clouds in atmosphere}
\label{sec:enstcon}
To find the placement of the enstatite cloud in our simulated atmosphere, we used the condensation curves from figure 4 of \citet{sea00}, which suggest that in our temperature pressure profile, enstatite condenses at approximately $4.1\times 10^8$ cm, or about halfway up the atmosphere. We chose to have the enstatite cloud continue through two pressure scale heights, similar to Jupiter's clouds, which places the upper boundary of the cloud at approximately $5.8\times 10^8$ cm, or 70\% the height of the atmosphere. 

Using the same techniques as for enstatite, we find that the iron should condense at $3.4\times 10^8$ cm, or about 42\% up the atmosphere. We again chose to have the iron cloud continue through two pressure scale heights, which puts the upper boundary of the iron cloud at approximately $5.1\times 10^8$ cm, or 62\% the height of the atmosphere.

The clouds we simulate are fractal, largely without sharp edges, and they follow the same pressure profile, meaning they will usually be optically thicker at the bottom than at the top.

\subsubsection{Opacity and scattering properties of hydrogen, enstatite, and iron}
Hydrogen opacity is calculated from the cross section calculated by \cite{dal62},
\begin{equation}
\displaystyle
\sigma(\lambda)=\frac{8.14\times10^{-13}}{\lambda^4}+\frac{1.28\times10^{-6}}{\lambda^6}+\frac{1.61}{\lambda^8}
\end{equation}
where the wavelength $\lambda$ is in \AA$ $ and the cross section is in $\rm{cm^2}$. Further, using the relationship $\sigma n= \kappa\rho$
where $n$ is the number density, we find the opacity of molecular hydrogen at 500 nm is $\kappa=4.14\times10^{-4} {\rm cm}^2\,{\rm g}^{-1}$. Knowing the density and opacity of molecular hydrogen, and assuming Rayleigh scattering, allows us to include molecular hydrogen in our simulations. Though we generally assume hydrogen acts only as a scatterer (an albedo of 1), we assigned hydrogen an albedo of 0.999, so that photons that passed through the enstatite and iron clouds would not scatter around the optically thick hydrogen bottom layer indefinitely, but instead would eventually be absorbed. This is a reasonable approximation because it is unlikely that any atmosphere will consist of pure hydrogen and will likely contain absorbing contaminants.

The opacity and albedo of enstatite and iron are calculated using a Mie scattering code \citep{boh83}. For enstatite at 500nm the optical constants $n$ and $k$ are 1.577 and $2.1\times 10^{-5}$, respectively \citep{dor95} . Using these constants and assuming the enstatite is in the form of 5 $\mu m$ diameter spheres, we calculate the opacity of enstatite to be $1.12 \times 10^4 {\rm{cm}^2}{\rm{g^{-1}}}$ and the albedo $a_{\rm{MgSiO}_3}=0.999$. Using the optical constants of iron \citep{ord85}, $n=2.74$ and $k=2.88$, along with the assumption that the condensates are 5 $\mu m$ in diameter, allows us to use Mie theory to calculate the optical parameters of our single scattering iron particles at 500 nm \citep{boh83}. We find that the opacity of iron is $4.03\times10^3 {\rm{cm}^2}{\rm{g^{-1}}}$ and that the albedo is $a_{\rm{Fe}}=0.685$. 

\begin{figure}
  \includegraphics[angle=0,width=.5\textwidth]{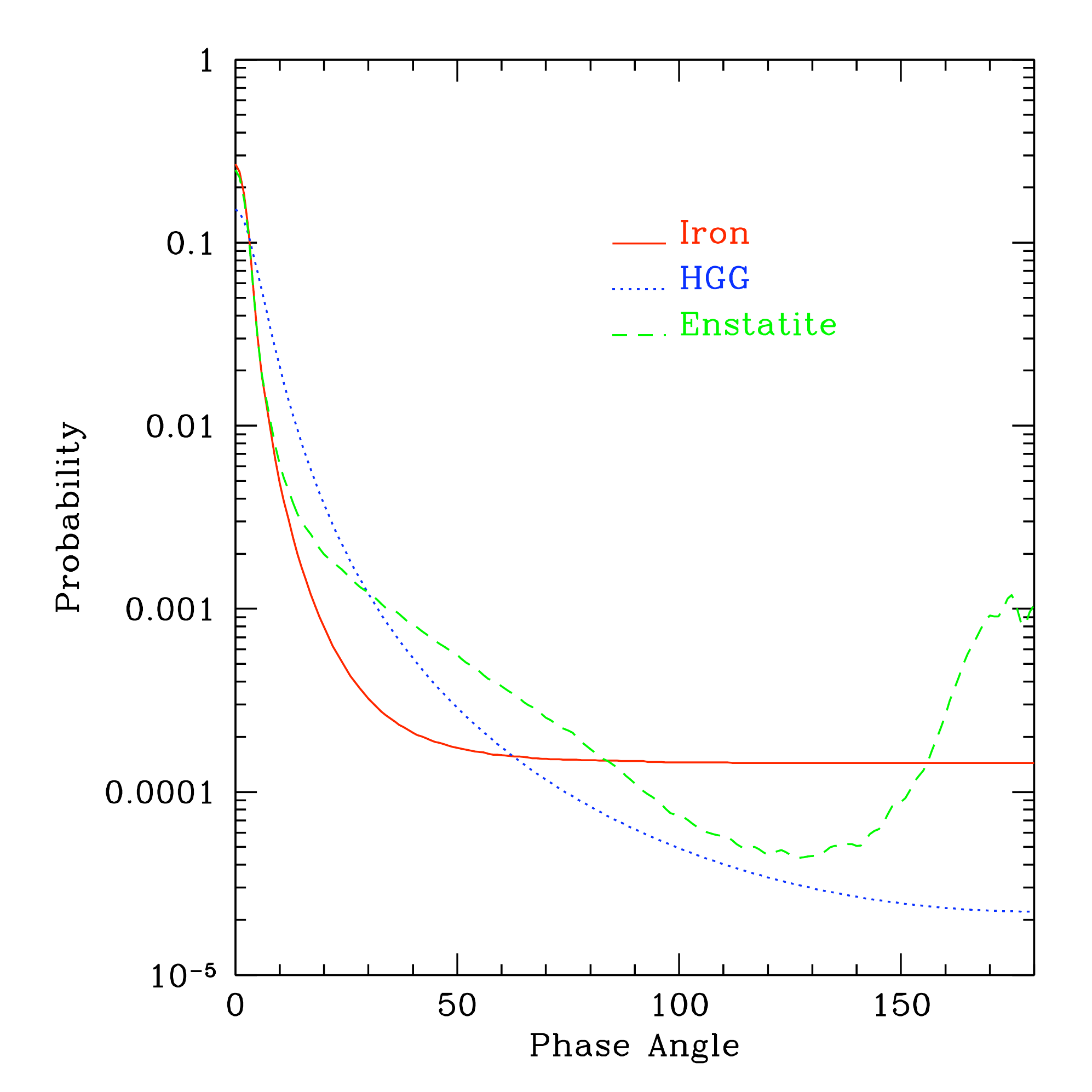}
  \caption{Henyey Greenstein versus Mie scattering phase functions for iron and enstatite. The Mie phase functions are calculated for five micron radii enstatite and iron grains. The enstatite has a slight back scatter around 180 degrees. The Henyey Greenstein g=0.9 phase is similar to the enstatite phase, but has no back scatter. The g=0.9 phase also underestimates the iron phase at 180. We opted to use the Mie phase functions in this simulation of HD 209458b.}
  \label{fig:hg}
\end{figure}

The scattering phase functions for enstatite and iron are calculated from Mie theory and tabulated for the Monte Carlo scattering code. Using pre-computed tabulated phase functions allows a wide variety of phase functions to be incorporated into our scattering codes, and for enstatite allows us to have a slight back-scattering component not possible with a Henyey-Greenstein phase function (see figure \ref{fig:hg}).

\begin{figure}
\includegraphics[width=.45\textwidth]{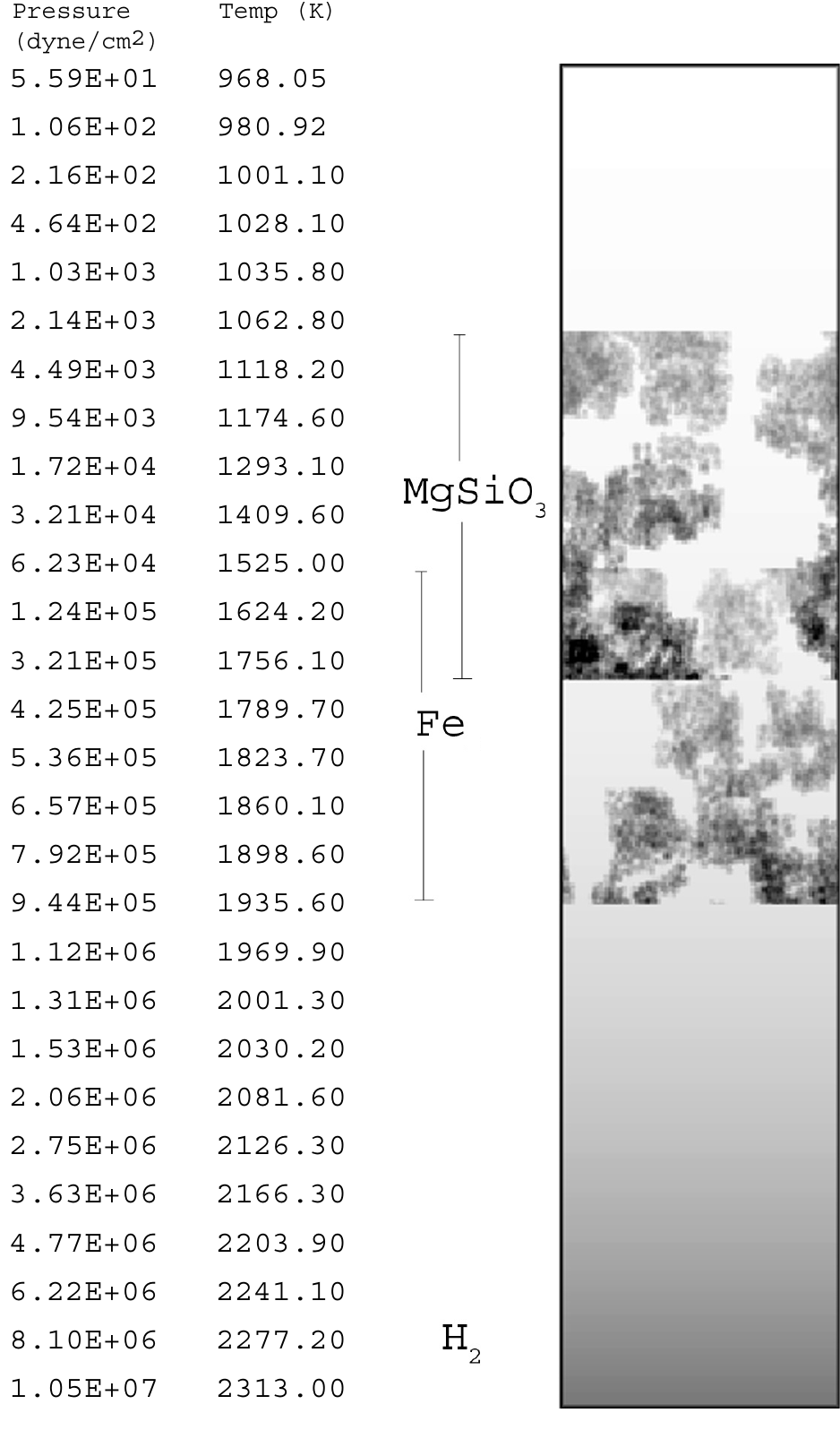}
\caption{Density$\times$opacity schematic of the atmospheric simulation for HD 209458b. This is a column summation for all $y$. Our entire simulation is approximately 8000 km in height, with the temperature and pressures noted at the left. The dark regions of the schematic are optically thickest while the white regions are optically thin. Molecular hydrogen is throughout the exponential density atmosphere. Enstatite clouds condense highest in the atmosphere, and overlap with the iron which condenses slightly below them. Certain features have been exaggerated in this schematic, such as the opacity due to hydrogen, and the tenuity of the enstatite and iron clouds, in order to see the features more clearly.}
\label{fig:grad}
\end{figure}

\subsection{Results}
\label{sec:209res}
Before reporting our results, we should note that our fiducial 1D model is the only self-consistent atmospheric structure model we use. In our 3D scattered light simulations, we freely move the condensate clouds, adjust the opacities (by changing the mass), and change the absorption of the gas. We do this primarily because we want to isolate the effects of our changes. We also do this because there does not yet exist a fully self-consistent 3D atmosphere model that accurately includes 3D radiation transfer effects, although there are moves to address this\citep{sho08}. Our variations can be equated to small changes in the modeled atmosphere. For instance, changing the height of the clouds can be equated to using a different temperature pressure profile. Changing the absorption of the gas assumes that a species other than molecular hydrogen might be prevalent in the atmosphere. Using a different albedo for a condensate tests the effects of an impure (or altogether different) condensate in the atmosphere. By changing these individual input parameters, we no longer create a self-consistent model, but we are able to isolate the effect of our variations. We feel that with this first 3D modeling of scattered light from an extrasolar planetary atmosphere, isolating the effect of our variations is more important than a self-consistent simulation.

\subsubsection{Fiducial 3D model}

\begin{figure}
  \includegraphics[angle=0,width=.5\textwidth]{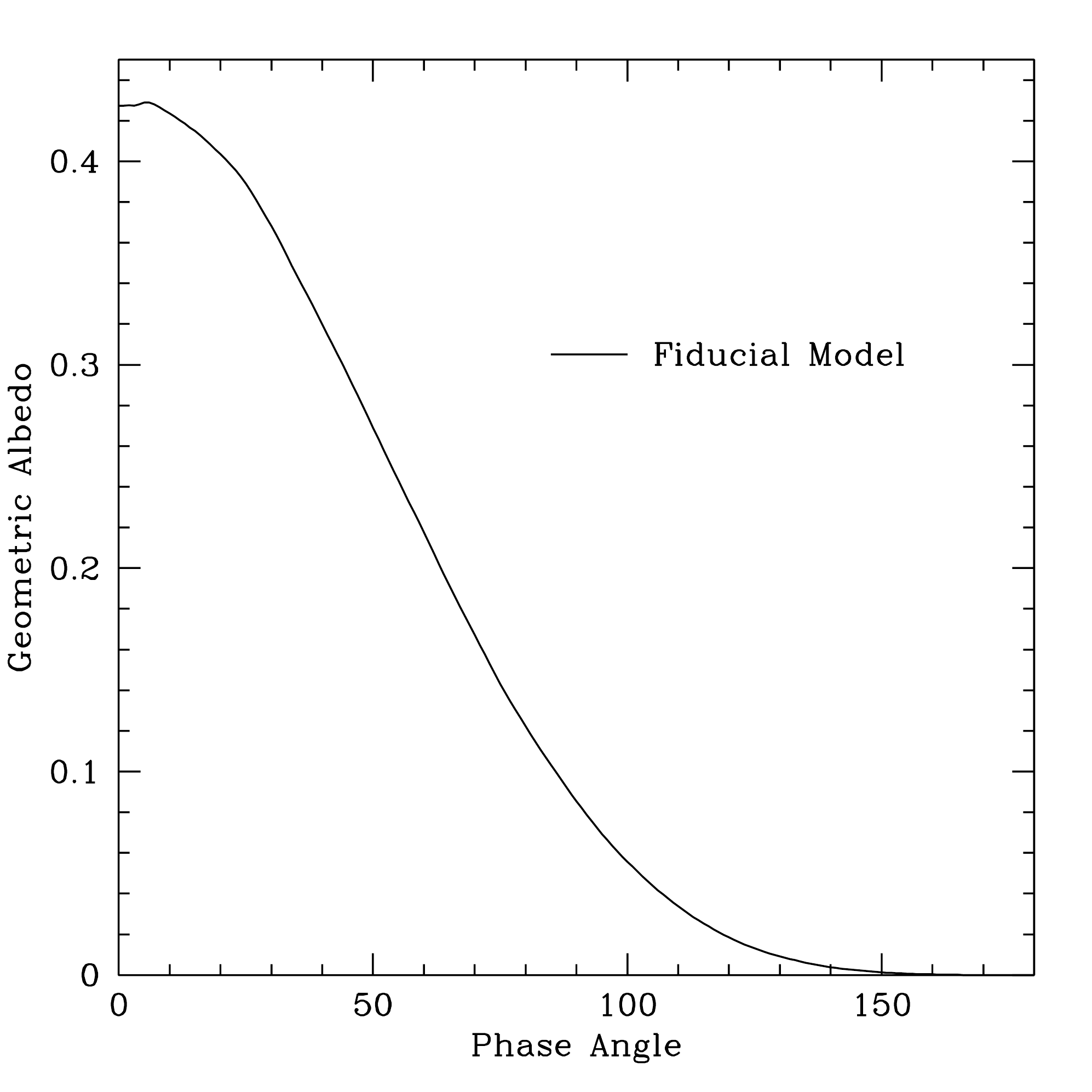}
  \caption{Geometric albedo phase curve of our fiducial model, including a two pressure scale height enstatite cloud and a similar, but lower, iron cloud in the atmosphere, with Rayleigh scattering hydrogen gas throughout.}
  \label{fig:normal}
\end{figure}

Using the parameters computed above for the 1D density structure, we have created a fiducial 3D model of HD 209458b which generates a geometric albedo $A_g =0.42$ at 500 nm. The atmospheric scattered light simulation consists of three scatterers: Rayleigh scattering hydrogen gas, Mie scattering five $\mu$m enstatite, and Mie scattering five $\mu$m iron. The pressure and temperature of HD209458b are provided by Seager's 1D calculation from which the density is computed. We assume a 10\% condensation rate, $c$, for both enstatite and iron. We convert this 1D model to 3D as described above, assume a porosity (as described in \S\ \ref{sec:atmgeo}) of 45\%, and we assume that the particulates will condense according to their condensation curves and will extend up two pressure scale heights \citep{sea05}. Our 3D radiation transfer simulation models the outermost $\sim$ 8000 km of the atmosphere. It is against this fiducial model (see figure \ref{fig:grad}) that we will compare our other simulations. With these fiducial parameters, the resultant geometric albedo of HD 209458b is $A_g\sim 0.42$, as seen in figure \ref{fig:normal}.

\subsubsection{Abundances and condensation}

To determine the extent of condensed enstatite and iron in the atmosphere, we use a single abundance parameter that encompasses both condensation and abundance. For instance, in our model, an atmosphere with 10x solar abundance and 10\% condensation is treated the same as a 1x solar abundance with a 100\% condensation.

For our fiducial model, we assumed that 10\% of the enstatite and iron in the atmosphere would condense into 5 $\mu m$ grains. The condensation of enstatite and iron could be affected by the purity of the metals in the atmosphere, as well as the turbulence.

Our fiducial model initially assumes a solar abundance of the constituents of enstatite and iron. However, the planetary system as a whole could have a higher abundance than solar, as HD 209458 has a higher abundance than solar \citep{sch}. Additionally, HD 209458b could have a higher or lower abundance of enstatite or iron than solar, as Jupiter and Saturn have some metals at differing concentration rates \cite{gui99}. Finally, it is also possible that condensates could form at higher regions in the atmosphere and rain down into the cloud deck, creating locally higher or lower concentrations of enstatite and iron than the planet as a whole \cite{ack01}.

Because our model only considers this single parameter which encompasses both condensation and abundance, we vary our enstatite and iron levels widely to encompass both low abundances and condensation rates, as well as high abundances and high condensation rates. We vary this parameter from 0.001 to 10 times solar abundance.

\begin{figure}
  \includegraphics[angle=0,width=.5\textwidth]{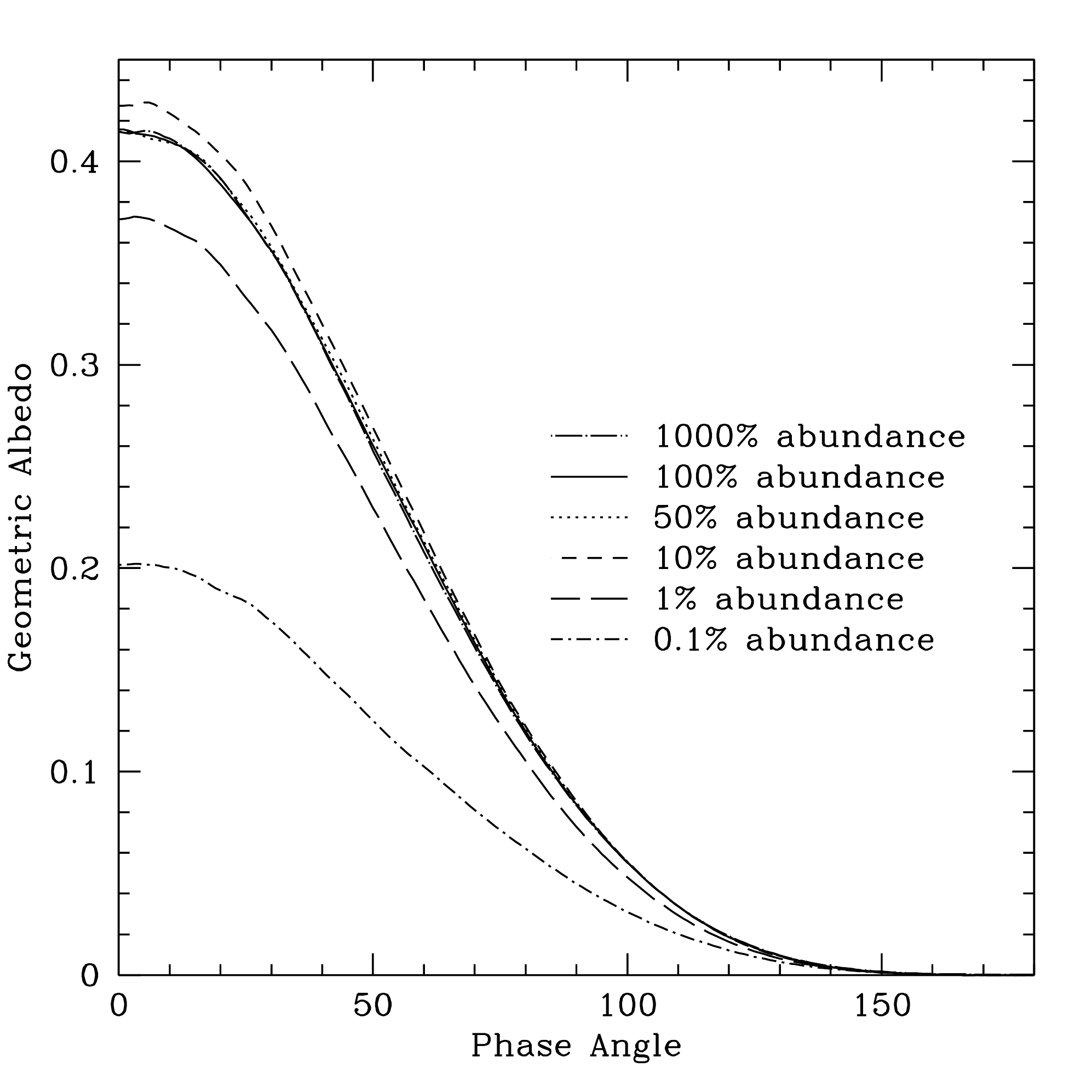}
  \caption{Geometric albedo phase curve of varying abundance rates for the iron and enstatite clouds. We vary some values above 100\% because HD 209458b may be more metal-rich than our sun, resulting in more enstatite in the cloud.}
  \label{fig:condense}
\end{figure}

Figure \ref{fig:condense} demonstrates that the abundance of enstatite and iron in the cloud only has a small impact on the geometric albedo, when it is above a minimum amount, around 1\% of solar abundances. Similar to our earlier, more general, models (see \S\ \ref{sec:simres}), larger opacities combined with a fractal cloud lower the overall reflectivity. This leads to a 100\% abundance producing a slightly less reflective geometric albedo than our 10\% fiducial model. This is due to the light entering the cloud at optically thin regions but not escaping because it is surrounded by optically thick regions. However, even the 1000\% abundance model shows only modest differences in the geometric albedo from the fiducial model, within the noise limits of randomly changing the cloud structure but maintaining the same porosity.

\subsubsection{Porosity}
Our fiducial scattered light model assumed that the cloud porosity is 45\% and we have compared this with models with porosities of 80\% and  2\%. It should be noted that when we consider percent porosity, we are only considering the grid cells containing the condensate cloud, the hydrogen is considered to be smoothly distributed with the exponential density described above. In all of our porous models the clouds have $f_{\rm{smooth}}=0$.  

\begin{figure}
  \includegraphics[angle=0,width=.5\textwidth]{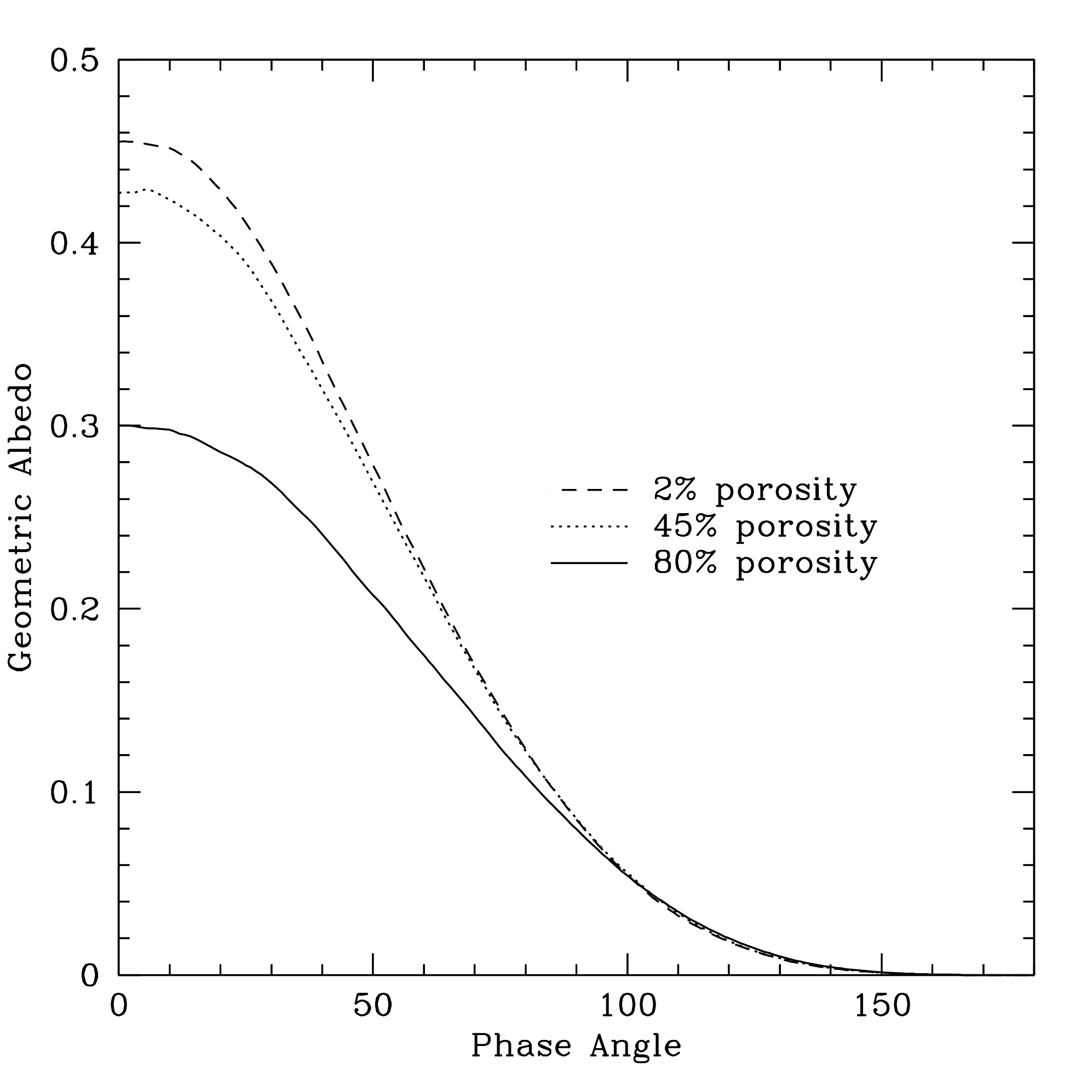}
  \caption{Geometric albedo phase curve of varying porosity rates. The nearly smooth model (2\% porosity) returns a geometric albedo only slightly better than our fiducial model of 45\% porosity. The more tenuous 80\% porosity model gives a much less reflective planet.}
  \label{fig:porous}
\end{figure}

As expected from our general investigations, figure \ref{fig:porous} shows that cloud porosity is a significant factor in determining the overall reflectivity of HD209458b. If incident starlight is able to travel deeply into a highly porous cloud, through the optically thin regions, then that light is often not able to emerge from the atmosphere, and results in a lower geometric albedo. The porosity is even more important in the specific situation of HD 209458b than it is in our general cases in \S\ \ref{sec:simres}. This is because of the relatively low albedo of iron ($a_{\rm{Fe}}=0.685$) compared to enstatite ($a_{\rm{MgSiO_3}}=0.999$). In general the enstatite cloud will condense at a higher altitude than the iron cloud, meaning that in a smooth cloud deck the photons will mainly interact with the optically thick enstatite and never reach the lower iron cloud. However, a porous enstatite cloud will allow starlight to penetrate deeper into the atmosphere, where it may be absorbed by the low albedo iron. By comparison, in our general investigations changing the porosity from 45\% to 80\% reduced the geometric albedo by about 30\%. The same change in the HD 209458b atmosphere reduced the geometric albedo by 40\%.

\subsubsection{Cloud height}
The height at which the enstatite and iron clouds form may be varied and this is equivalent to exploring situations in which the atmospheric clouds have not settled as well as predicted from our 1D temperature-pressure model. Cloud heights may differ due to a very turbulent atmosphere, or because there is a higher abundance of iron than we suppose. Additionally, changing the cloud height is a way we can test different temperature pressure profiles. We are aware that this is not self-consistent, in that we are placing the clouds at temperatures and pressures where they would not occur, but this still gives us good first approximations as to what would happen with completely different parameters. We tested cases in which the iron and the enstatite occur at the same cloud height, and in the unlikely situation of iron occurring at the enstatite height in the atmosphere, and enstatite occurring at the iron height. 

\begin{figure}
  \includegraphics[angle=0,width=.5\textwidth]{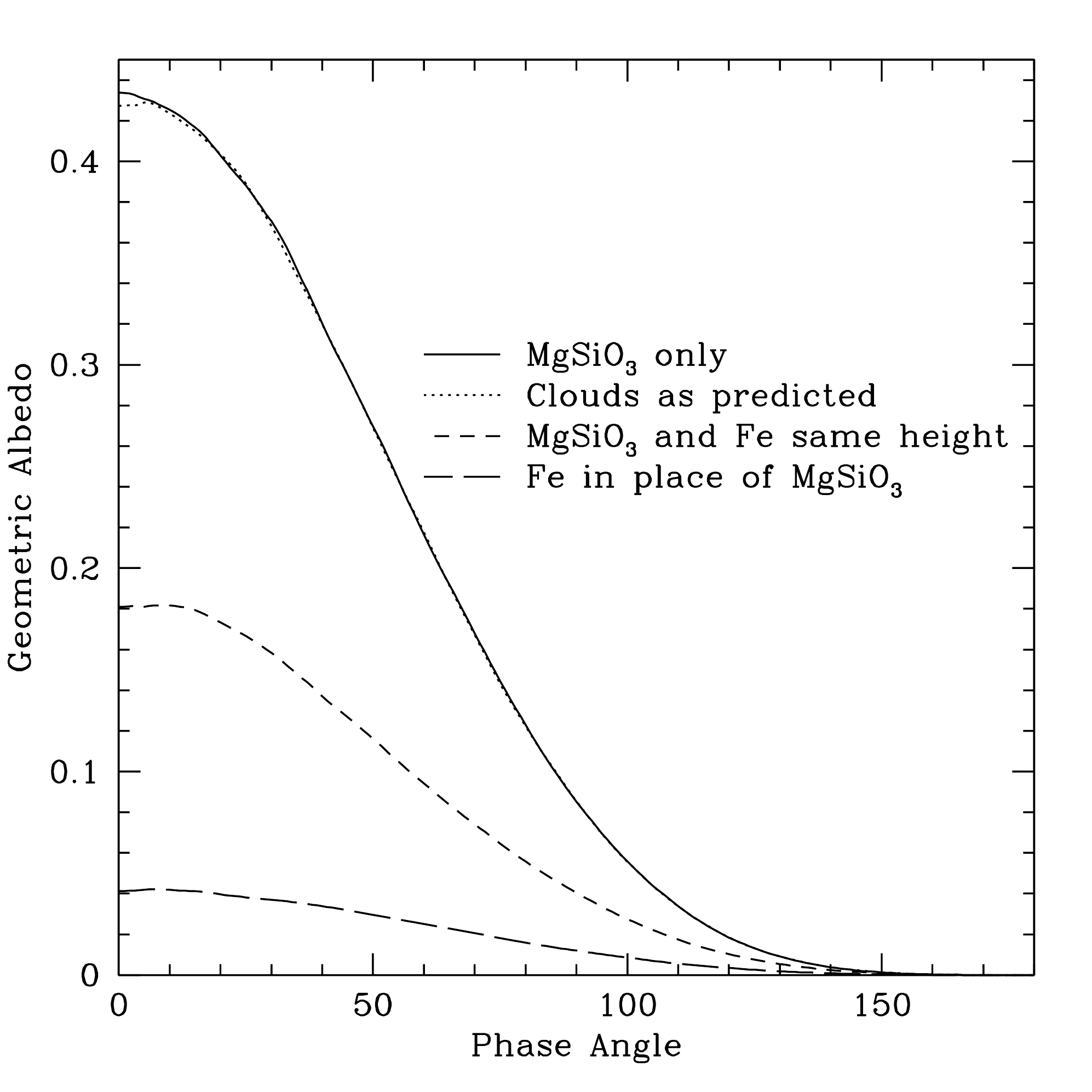}
  \caption{Geometric albedo phase curve of cloud heights. Our fiducial (predicted) model puts the enstatite cloud highest in the atmosphere, with the iron below it. If we instead mix the two clouds, and put both iron and enstatite at the same height in the atmosphere, the geometric albedo is reduced. In the situation where we switch the placement of iron and enstatite, with iron on top, the geometric albedo is significantly reduced, to only 10\% the value of our fiducial model. }
  \label{fig:cloud}
\end{figure}

Figure \ref{fig:cloud} shows the height of the condensate clouds hugely affects the geometric albedo of the planet, for much the same reason as the porosity rate affects it. In this case, it is mainly the height of the iron cloud that makes the difference. Its low albedo can sharply decrease the overall reflectivity of a planet, producing an albedo only 14\% as reflective as the fiducial model. 

\subsubsection{Gas absorption}
To simulate additional absorptive constituents in the hydrogen gas component of the model atmosphere we ran scattered light simulations in which the gas albedo is lowered from 0.999 to 0.5. This is a possibility if TiO, C$\rm{H}_4$, or Na molecules are in the atmosphere, because they are much more absorptive than hydrogen. The simulations also explored the effects on the planet's geometric albedo of raising and lowering the height of the enstatite cloud.

\begin{figure}
  \includegraphics[angle=0,width=.5\textwidth]{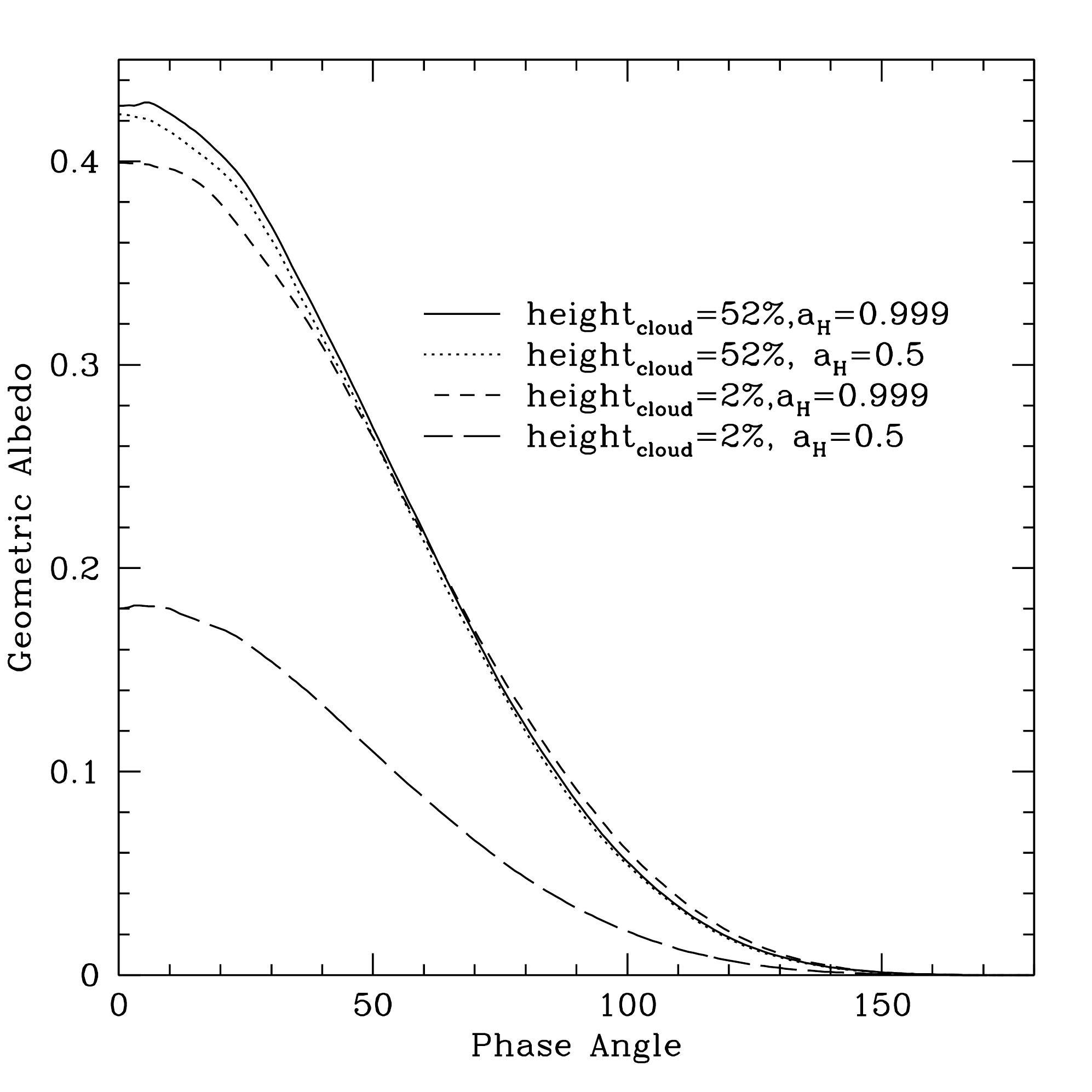}
  \caption{Geometric albedo phase curve of different absorption for the gas, and different heights of the enstatite cloud. Changing only the gas absorption does not affect the geometric albedo much, nor does making the enstatite cloud lower. However, if we make the gas more absorptive and lower the cloud, then the light is absorbed by the gas before it reaches the enstatite cloud, leading to a lower geometric albedo.  }
  \label{fig:atmoalbedo}
\end{figure}

Figure \ref{fig:atmoalbedo} shows there is very little difference between the case where the enstatite clouds were high in the atmosphere and the gas albedo was varied. The opacity of the molecular hydrogen is too small to make much of a difference in the geometric albedo when most photons instead interact with the much more opaque and higher enstatite and iron. However, the case changes when the clouds are moved to the bottom of the atmosphere. Even though the clouds are much more opaque than the hydrogen, the photons still interact and are absorbed if the cloud is around 2\% of the height of the atmosphere instead of 52\% high. Merely lowering the cloud or lowering the molecular hydrogen albedo will not change the geometric albedo, but the two changes together cut the geometric albedo by more than half.

\subsubsection{Enstatite albedo}
\label{sec:ensalb}
\begin{figure}
  \includegraphics[angle=0,width=.5\textwidth]{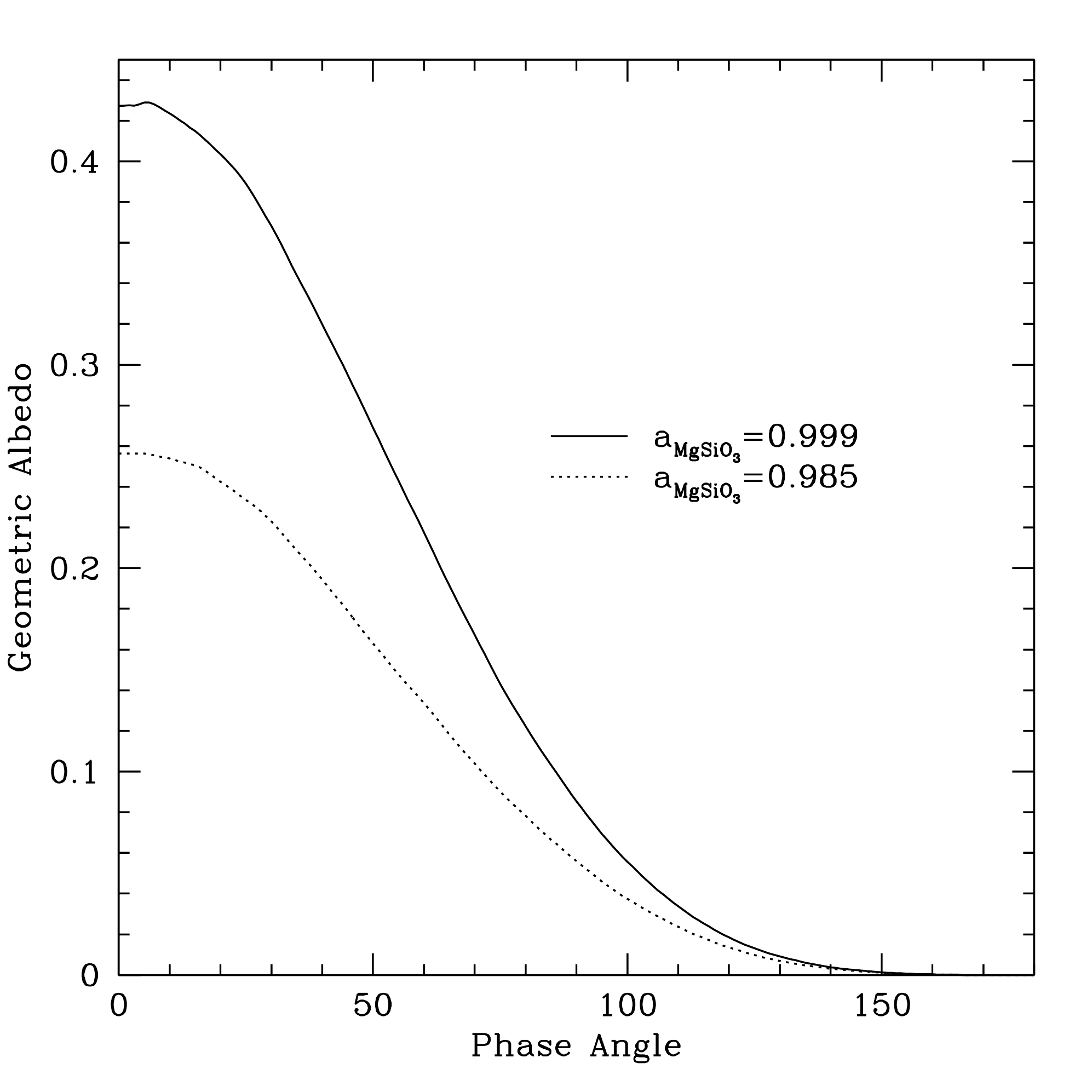}
  \caption{Geometric albedo phase curve of different single scattering albedos for enstatite. The relatively small change in the single scattering albedo, from 0.999 to 0.985 has a large impact on the geometric albedo of the planet. This change in single scattering albedo could result from more amorphous than glassy enstatite in the atmosphere. Note that we kept the same phase function for enstatite, and only changed the albedo.}
  \label{fig:enstalbedo}
\end{figure}

By using a slightly lower value for the single scattering albedo of enstatite, our models produce a geometric albedo of only $A_g = 0.25$. Using Mie theory and optical constants for glassy enstatite from \citet{dor95}, we calculate a single scattering albedo for enstatite of $a_{\rm{MgSiO_3}}=0.999$. However, optical constants by \citet{sco96} differ slightly, and measure amorphous enstatite. These values are used by \citet{mar99} to develop their extrasolar giant planet atmospheric model. Using the values of \citet{sco96} results in an albedo of enstatite of $a_{\rm{MgSiO_3}}=0.985$. This minute change contributes significantly to the geometric albedo, because the phase function of enstatite throws most photons forward, so most photons penetrate deep into the atmosphere and scatter many times before coming out. Because the photons scatter so many times, a small change in the albedo contributes significantly to the geometric albedo, as we show in figure \ref{fig:enstalbedo}. Comparing glassy and amorphous enstatite is outside the scope of this work, but it is sufficient to say that the constituents of extrasolar planetary atmospheres are not known well-enough to favour one or the other. It should be noted that though we used the adjusted optical constants to compute the second albedo, we did not recompute the phase function of the enstatite based on the new constants. This was to isolate the change in albedo and we see that what appears to be a minor albedo difference, between amorphous and glassy enstatite, can nearly halve the geometric albedo of the planet.

\subsection{Discussion of HD 209458b Model Atmosphere}
\label{sec:209dis}
For HD 209458b to have a high geometric albedo, the enstatite cloud condensate must be high in the atmosphere, smooth, and pure. At the temperatures and pressures present in the atmosphere of HD 209458b, there are very few condensates which can make a positive impact on the geometric albedo of the planet \citep{sea00}. According to the condensation curves, enstatite forms high in the atmosphere and has a single-scattering albedo high enough to give the planet a high geometric albedo. 

However, the geometric albedo of HD 209458b has a 3$\sigma$ upper limit of 0.17 in the visible wavelengths, as determined by the \emph{MOST} satellite \citep{row06,row07}. This upper limit on the geometric albedo seriously constrains the nature of the enstatite in the atmosphere of HD 209458b. The enstatite cannot be high, smooth and pure as we speculated in our fiducial model, because if it were, then the reflected starlight would have been detected by \emph{MOST}.

We find several variations of our models which depress the geometric albedo of HD 209458b. Lowering the abundance (comprised of both condensation and abundance compared to solar) from 10\% to 0.1\% cuts the geometric albedo by more than one half. Very porous clouds lower the albedo by one quarter. Significantly lowering the enstatite cloud and raising the gas absorption together more than halve the geometric albedo, though each change on its own does not affect the geometric albedo. Lowering the single scattering albedo of enstatite to 0.985 from 0.999 reduces the geometric albedo $A_g =  0.25$ compared to our fiducial model value $A_g = 0.42$. Finally, mixing the high-albedo enstatite with the low-albedo iron more than halves the geometric albedo, and if iron is higher in the atmosphere, then the geometric albedo is only $A_g = 0.04$.

Though a 0.1\% abundance of enstatite results in a low geometric albedo for the planet, we are unconvinced that a low abundance of enstatite is what is causing the low albedo of HD 209458b. First, the metallicity of HD 209458 is higher than solar metallicity, suggesting that the enstatite in the system could be higher than solar \citep{sch}. Second, we notice the enhanced metallicity in the atmospheres of our own gas giant planets, Jupiter and Saturn, compared to solar, and consider it reasonable that extrasolar giant planets will also exhibit an enhanced metallicity compared to the parent star. Finally, we think the cloud condensation can promote a locally enhanced abundance at the condensation height of enstatite in the atmosphere.

If clouds are very porous, the geometric albedo is reduced, but not enough to explain the low geometric albedo. In our three test simulations, the most porous case with 80\% porosity still gives a geometric albedo $A_g = 0.3$, above the upper limit of the \emph{MOST} data.

If the cloud is much lower in the atmosphere than the condensation curves suggest, and the atmosphere is filled with more absorptive Rayleigh scatterers, then the geometric albedo is reduced to $A_g\sim 0.18$, still above the upper limit imposed by \emph{MOST}. This is a reasonable situation if ions or other species expected in the atmosphere are highly absorptive. However, as mentioned before, the opacities of the condensates are much higher than the opacities of the Rayliegh scatterers, so the condensates must be very low in the atmosphere, and the Rayleigh scatterers very absorptive to generate a low geometric albedo.

The nature of the enstatite, either glassy or amorphous, significantly affects the geometric albedos, because it changes the single scattering albedo from 0.999 to 0.985. This small change reduces the geometric albedo from $A_g = 0.42$ to 0.26, above the geometric albedo limit of \emph{MOST}. We do find it hard to believe that all the enstatite condensed in the atmosphere is pure, glassy enstatite. Other species are present in the atmosphere, even some other condensates, so it is a reasonable expectation that some of the reduction in geometric albedo comes from amorphous and impure enstatite depressing the single scattering albedo of the condensate, but obviously, just the change to amorphous enstatite is not sufficient, because the geometric albedo remains above our upper limit.

The atmospheres of Jupiter and Saturn are convective and as a result mostly homogenous. The upper layers are not completely homogenous, due partly to condensation of clouds, but there is still diffusion of heavy elements throughout the atmosphere \citep{gui99}. In our simulations where the iron and the enstatite have mixed, the geometric albedo drops to less than half the value of our fiducial model. If iron is above the enstatite, the geometric albedo drops to $A_g = 0.04$. Several groups comment on the turbulence expected in the atmosphere of HD 209458b \citep{sho02,cho03}. We think that a turbulent atmosphere, with routine mixing of various condensates (and the resulting drop in the average single scattering albedo), is a significant contributor to the low geometric albedo. 

\section{Summary}
\label{sec:209sum}

The primary goal of this paper was to investigate the effects of 3D radiation transfer on the optical reflected light levels from extrasolar planetary atmospheres. We find that compared to smooth models, 3D atmospheres reflect less light because starlight can penetrate deep into the atmosphere, increasing its probability of being absorbed. As found in other 1D and 2D scattered light simulations, other atmospheric parameters serve to reduce the geometric albedos of extrasolar planets, most notably the scattering albedo of individual particles in the planetary atmosphere \citep{dlu74}. Indeed, we suspect that it is the albedo of particles in the atmosphere that results in the very low geometric albedo observed from HD209458b.

We realize that our model is not a perfect representation of an extrasolar atmosphere, and that other effects not explored here might contribute to a lower geometric albedo. Specifically, strong absorption of alkali metals sodium and potassium could lower the albedo throughout the visible spectrum (\cite{sud00} and \cite{sea00}). However, we have tried to isolate some effects, specifically from the 3d structure of the atmosphere, that might contribute to the overall albedo of extrasolar planets.

HD 209458b is a dark planet, with a geometric albedo $A_g < 0.17$. One possibility is that clouds are absent and the low
albedo is due to atomic or molecular absorption. To study the albedos in the presence of clouds, we investigated a fiducial 3D scattered light model which considered the densities, opacities, and albedos of hydrogen, iron, and enstatite, and their respective placements in the atmosphere. This fiducial model gives a geometric albedo $A_g > 0.4$, so we know that it does not accurately represent the atmosphere of HD 209458b. Upon varying parameters in our model we found significant drops in the geometric albedo. Parameters that strongly affected the geometric albedo include making the clouds more porous, lowering the condensation rate of enstatite, and assuming amorphous enstatite instead of glassy. Indeed, we were surprised by how easily the geometric albedo could be reduced with a relatively small change to the constituents of the atmosphere. However, we feel the most important parameter influencing the reflected light levels is the purity of enstatite and the assumption that it sits alone and above all the other condensates in the atmosphere. Models have suggested that HD 209458b has a turbulent atmosphere, and if clouds are indeed present we expect that mixing of enstatite with other, lower albedo condensates is what keeps the geometric albedo so low.

\section*{Acknowledgements}

Ben Hood would like to thank the Marshall Commission and the Department for 
Terrestrial Magnetism of the Carnegie Institution of Washington for financial 
support.

\bibliography{reference} 
\bibliographystyle{mn2e}

\end{document}